\documentclass[12pt]{article}

\usepackage{scicite}

\usepackage{times}

\usepackage{amsmath}
\usepackage{amsfonts}
\usepackage{amssymb}
\usepackage{graphicx}
\usepackage{caption}
\usepackage{siunitx}

\usepackage[letterpaper, textwidth=17cm]{geometry}

\newenvironment{sciabstract}{%
\begin{quote} \bf}
{\end{quote}}

\title{Fundamental scaling limits and bandwidth shaping of frequency-modulated combs}

\author
{Mithun Roy,${}^{1\ast}$ Zhenyang Xiao,${}^{1}$ Sadhvikas Addamane,${}^{2}$ David Burghoff${}^{1}$\\
\\
\normalsize{${}^{1}$Department of Electrical and Computer Engineering, University of Texas at Austin,}\\
\normalsize{Austin, Texas 78712, USA}\\
\normalsize{${}^{2}$Center for Integrated Nanotechnologies, Sandia National Laboratories,}\\
\normalsize{Albuquerque, New Mexico 87185, USA}\\
\\
\normalsize{$^\ast$To whom correspondence should be addressed; E-mail:  mithunroy177@utexas.edu.}
}

\date{}

\begin{document} 


\baselineskip24pt


\maketitle

\begin{sciabstract}
    Frequency-modulated (FM) combs based on active cavities like quantum cascade lasers have recently emerged as promising light sources in many spectral regions. Unlike passive modelocking, which uses amplitude modulation to generate amplitude modulation, FM combs use phase modulation to generate phase modulation. They can therefore be regarded as a phase-domain version of passive modelocking. However, while the ultimate scaling laws of passive modelocking have long been known---Haus showed in 1975 that pulses have a bandwidth proportional to effective gain bandwidth---the limits of FM combs have been much less clear. Here, we show that FM combs are governed by the same fundamental limits, producing combs whose bandwidths are linear in the effective gain bandwidth. Not only do we show theoretically that the diffusive effect of gain curvature limits comb bandwidth, we also show experimentally how this limit can be increased. By adding carefully designed resonant-loss structures that are evanescently coupled to the cavity of a terahertz laser, we reduce the curvature and increase the effective gain bandwidth of the laser, demonstrating bandwidth enhancement. Our results give a new degree of freedom for the creation of active chip-scale combs and can be applied to a wide array of cavity geometries.
\end{sciabstract}



\section{\label{intro}Introduction}

Self frequency-modulated (FM) combs are a class of optical comb that produce not pulses, but constant-intensity chirped waves. Although hints of their existence were observed for decades \cite{lauPassiveActiveMode1985,paiellaSelfModeLockingQuantumCascade2000,heckObservationQswitchingModelocking2007,heckPassivelyModeLocked102009}, the first conclusive identification of linearly-chirped operation was observed in quantum cascade lasers (QCLs) in 2018 \cite{singletonEvidenceLinearChirp2018,hugiMidinfraredFrequencyComb2012,burghoffTerahertzLaserFrequency2014,henryTemporalCharacteristicsQuantum2018}. Later, this same mode of operation was found to occur in a wide variety of semiconductor lasers, ranging from diode lasers to quantum dot lasers to VCSELs \cite{sterczewskiFrequencymodulatedDiodeLaser2020,daySimpleSinglesectionDiode2020,dongBroadbandQuantumdotFrequencymodulated2023,krisoFrequencymodulatedCombVECSEL2021,hillbrandInPhaseAntiPhaseSynchronization2020}. The phenomenon is surprisingly robust and general. FM comb generation in QCLs is of interest as they can operate both in the mid-infrared (mid-IR) and terahertz (THz) regions, while their compact and broadband nature makes them suitable for many applications, such as spectroscopy \cite{villaresDualcombSpectroscopyBased2014,yangTerahertzMultiheterodyneSpectroscopy2016,burghoffComputationalMultiheterodyneSpectroscopy2016,burghoffGeneralizedMethodComputational2019}, radiometry \cite{benirschkeFrequencyCombPtychoscopy2021}, metrology, and quantum information science \cite{gabbrielliIntensityCorrelationsQuantum2022}. However, the short gain recovery time of QCLs make passive modelocking challenging.


Interestingly, this mode of operation was not originally designed and happened somewhat spontaneously in Fabry-P\'erot (FP) cavities \cite{hugiMidinfraredFrequencyComb2012,khurginCoherentFrequencyCombs2014,burghoffTerahertzLaserFrequency2014}. Furthermore, it had not been theoretically predicted. The effect was originally believed to relate to a large third-order nonlinearity specific to intersubband systems, but the surprising generality and robustness indicated that that was not the case. Subsequent theoretical work established that FM comb generation was a broader emergent phenomenon that can arise in any laser due to the motion of the gain grating \cite{burghoffUnravelingOriginFrequency2020}. Briefly, gain saturation combines with the asymmetry in the field at the facet of a cavity to create an effective quasi-$\chi^{(3)}$ nonlinearity \cite{opacakTheoryFrequencyModulatedCombs2019,burghoffUnravelingOriginFrequency2020}. This nonlinearity creates a phase modulation dependent on the phase of the field itself, and this causes the system's dynamics to be governed by a phase-driven nonlinear Schrodinger equation \cite{burghoffUnravelingOriginFrequency2020}. The analytical solution is a chirped Gaussian, and so the natural state of these combs is 
an FM mode of operation, where the frequency is strongly modulated but the amplitude is not. 
FM-like states in different laser systems have been proven using a variety of techniques, primarily SWIFTS \cite{burghoffEvaluatingCoherenceTimedomain2015,hanSensitivitySWIFTSpectroscopy2020,singletonEvidenceLinearChirp2018} but also FACE \cite{cappelliRetrievalPhaseRelation2019}, upconversion sampling \cite{taschlerFemtosecondPulsesMidinfrared2021,taschlerAsynchronousUpconversionSampling2023}, fast photodetection with tunable filters \cite{heckObservationQswitchingModelocking2007,heckPassivelyModeLocked102009}, and stepped heterodyne \cite{dongBroadbandQuantumdotFrequencymodulated2023}. They mostly arise in semiconductor lasers because semiconductor lasers have high mirror losses and spatially extended cavities.

\begin{figure}
\centering	\includegraphics{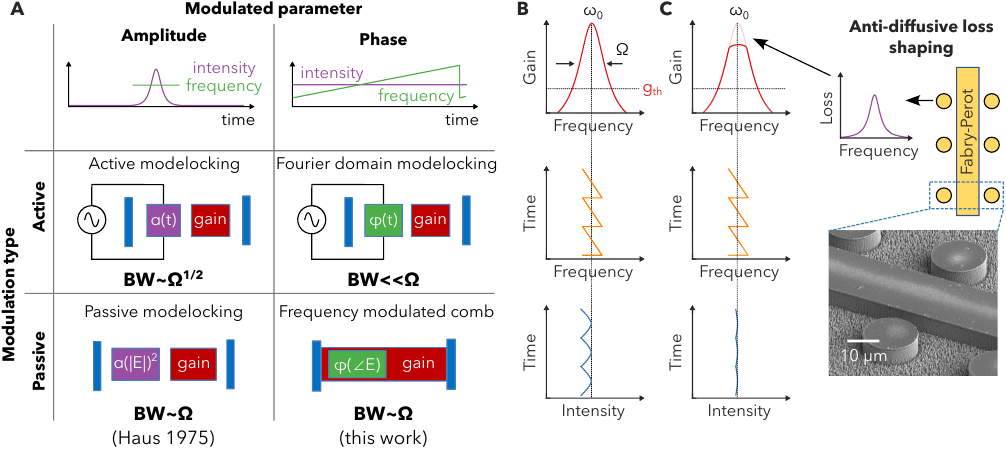}
\caption{\textbf{Fundamental bandwidth limits and gain curvature--stability interplay.}  \textbf{(A)} Overall landscape of laser-based combs. Traditional modelocking produces pulses by  amplitude modulation, while FDML and FM combs rely on phase modulation. Passive self-modulation produces broader bandwidths than external active modulation, and we show here that FM combs have the same linear dependence as passively modelocked lasers.
\textbf{(B)} Effect of gain curvature on FM comb formation. For an FM comb with large gain curvature, frequency modulation gives rise to intensity modulation that destabilizes the comb. \textbf{(C)} By introducing anti-diffusive loss shaping, in which resonant structures near the cavity reduce the gain curvature, unwanted amplitude fluctuations are suppressed.}
\label{fig:concept}
\end{figure}

Within the broader context of laser-based combs, FM combs can perhaps be understood as a generalization of the concept of modelocking (Fig. \ref{fig:concept}A). For many years, combs were primarily formed using \textit{amplitude} modulation. The first type of modelocking developed, active modelocking, generated amplitude modulation using an active loss modulator. Later, this active modulator was replaced with a passive saturable absorber, which allowed for the creation of shorter pulses by allowing them to modulate their own amplitudes. Eventually, techniques were developed that allowed for the creation of broadband combs based on \textit{phase} (or frequency) modulation. For example, Fourier domain modelocking (FDML) \cite{huberFourierDomainMode2006a} forms combs by actively tuning an intracavity filter synchronously with the mode spacing of the cavity. FM combs can be considered a passive equivalent of FDML, wherein the field's own phase generates additional phase modulation.

However, the ultimate bandwidth limits of FM modes of operation are not well understood. It is known that FM combs require low but nonzero dispersion to maintain stability. Nevertheless, extensive efforts in dispersion engineering have not yielded FM combs that fully exploit the available gain bandwidth. In contrast, the bandwidth limits for modelocked pulses are well-known and are described by the Haus theory \cite{Hausatheoryofforced,Haustheoryoffast}: for a gain bandwidth $\Omega$ the comb bandwidth is proportional to $\sqrt{\Omega}$ in the case of active modelocking and $\Omega$ in the case of passive modelocking. In the case of FDML, the bandwidth is limited for chip-scale lasers as it would require the ability to broadly tune a filter at the repetition rate, making it more well-suited to fiber lasers. For chip-scale devices, the bandwidth would typically be significantly narrower than the gain bandwidth. The chief difficulty of describing FM combs is that their underlying dynamics are complex and arise from a spatially- and temporally-inhomogeneous gain.

In this work, we show that the ultimate bandwidth limit of linearly-chirped FM combs is the  gain bandwidth, showing that they can be considered the phase equivalent to passive modelocking. Using our mean-field theory (which generalizes the Lugiato-Lefever equation to describe active cavities with large intra-round trip changes \cite{burghoffUnravelingOriginFrequency2020}), we analytically and numerically show that the primary limiting factor for these combs is diffusion arising from the variation in the gain, also known as gain curvature. As in the case of passive modelocking, this gain curvature can be related to the effective gain bandwidth as $D_g\sim\Omega^{-2}$. In fact, without gain curvature the fundamental FM comb---referred to as an extendon, as it must extend throughout the cavity with low amplitude modulation---can have infinite bandwidth, simply by reducing the dispersion to zero. However, with gain curvature, even when the laser is far above threshold and has ample gain across a broad bandwidth, there is no guarantee that the FM comb will actually be able to utilize all this gain. As the instantaneous frequency detunes from the center of the gain peak, the instantaneous intensity falls in concert (Fig. \ref{fig:concept}B). Other frequencies will be able to lase in this window instead, and for large values of curvature, the competition between these two processes leads to chaotic multimode behavior. Alternatively, if the gain is low, the diffusive effect of gain curvature can instead cause the laser to produce continuous wave (CW) light.

Moreover, we show here how this bandwidth can be enhanced by introducing the concept of \textit{anti-diffusive loss shaping} to balance the effects of gain curvature. By adding carefully designed resonant-loss structures that are evanescently coupled to the cavity of a THz QCL and controlling their size and distance (Fig. \ref{fig:concept}C), we are able to flatten the effective gain of the laser and broaden the comb bandwidth. We show experimentally that this strategy can produce combs whose bandwidth is close to the absolute gain bandwidth limit of a medium---the ability to make a single-mode laser. We verify the coherence of our combs and demonstrate their FM characteristics using SWIFTS. The ability to engineer both dispersion and diffusion enables new exciting prospects for integrated combs, as the aspects of anti-diffusive loss and gain in integrated active cavities have not been well-explored and provide new degrees of freedom.

\section{\label{concept}Theory and Design Concepts}
Gain curvature is intrinsic to any laser. The gain peaks at the transition frequency and falls as the frequency is detuned from it, giving rise to a negative $\frac{\partial^2g}{\partial\omega^2}$. For a homogeneously broadened transition, this curvature is related to the dephasing time. FM-type combs form in semiconductor lasers (such as QCLs) due to the combination of gain saturation with the asymmetry in the field at the facets. An FM comb with gain curvature can most simply be described by a phase-driven nonlinear Schrodinger equation as \cite{burghoffUnravelingOriginFrequency2020}
\begin{align}\label{nlse}
	\frac{\partial E}{\partial t} = \frac{i}{2}\left(\beta-i\frac{D_g}{2}\right) \frac{\partial^2 E}{\partial z^2} + i\gamma |E|^2 \angle E E  - r (\left|E\right|^2-P_0),
\end{align}
where $\beta$ is dispersion, $D_g\equiv-\frac{\partial^2g}{\partial\omega^2}$ represents gain curvature, the middle term reflects the nonlinear phase potential, and the final term resists amplitude modulation.  The term containing gain curvature is similar to the familiar diffusion equation. For pulsed lasers, it primarily serves to broaden pulses, but for FM combs the effect is more subtle. As the solution is already highly chirped, the diffusive effect of gain curvature serves primarily to modulate the output, converting the natural phase modulation into amplitude modulation \cite{humbardAnalyticalTheoryFrequencymodulated2022}. Furthermore, for a homogeneously-broadened transition, gain curvature is inversely proportional to the gain bandwidth squared ($D_g\sim T_2^2\sim\Omega^{-2}$). 

\begin{figure}
\centering
\includegraphics[]{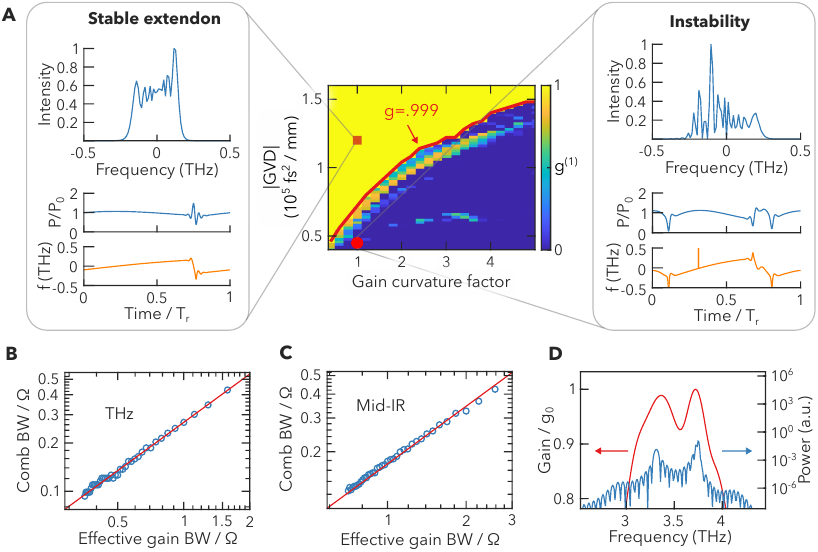}
	\caption{\textbf{Theoretical scaling laws.} \textbf{(A)} Comb coherence \cite{burghoffTerahertzLaserFrequency2014} as a function of GVD and gain curvature factor for a typical THz QCL.  For a dispersion higher than the optimum value, the intensity envelope is nearly constant. However, lower-than-optimum values result in chaotic amplitude fluctuations incoherent spectra. \textbf{(B, C)} Comb bandwidth as a function of effective gain bandwidth for a typical mid-IR and THz QCL, provided the optimum dispersion is chosen. These bandwidths are the maximum comb bandwidths that can be achieved for a given value of the effective gain bandwidth. \textbf{(D)} Effect of gain inhomogeneity on FM comb spectra, showing how a minor change in the gain can lead to an enormous spectral modulation. (Parameters in Supplementary Materials.)}
 \label{fig:theory}
\end{figure}

The core requirement that allows for stable extendon solutions to form is that the amplitude variation remains small.  To investigate the ultimate stability limits, we numerically used mean-field theory \cite{burghoffUnravelingOriginFrequency2020} to investigate how group velocity dispersion (GVD) and gain curvature affect the stability of mid-IR and THz QCL combs (Fig. \ref{fig:theory}A). We describe the comb as stable if the coherence is larger than 0.999, and find that the minimum value of GVD required to form a stable extendon depends on gain curvature. This minimum value, which we call the optimum dispersion, increases quadratically with gain curvature. Per the extendon theory, the comb bandwidth that can be achieved is inversely proportional to the effective GVD (which includes the effects of Kerr nonlinearity and linewidth enhancement \cite{humbardAnalyticalTheoryFrequencymodulated2022}). Therefore, the optimum dispersion is optimal in bandwidth as well. Even if the dispersion is chosen optimally, gain curvature limits the bandwidth.

By converting the gain curvature to effective gain bandwidth, as is done in the Haus theory, we are able to find the ultimate limits of FM combs. Figures \ref{fig:theory}B and C show the maximum achievable comb bandwidth as a function of the effective gain bandwidth (the gain bandwidth corresponding to the gain curvature factor), both for typical mid-IR and THz QCL parameters. As expected, the relationship is \textit{linear}, despite the fact that their gain recovery and dephasing times are significantly different. Not only did we find that this relationship holds over a wide parameter space (see Fig. S1B in Supplementary Materials), but also we found that by using perturbation theory and requiring that the amplitude variation be small compared to the CW amplitude, we can also derive this linear relationship \textit{analytically} (see ``Analytical relationship between comb and gain bandwidths" in Supplementary Materials). This general analytical relationship requires only modest assumptions---operation near threshold, approximately bilinear CW steady-state power function, and quasi-constant intensity.
For values of GVD exceeding the optimum dispersion, a nearly-constant-amplitude, linear chirp
forms (except at the jump point, where a pulse forms), and the spectrum is highly coherent (Fig. \ref{fig:theory}A left). On the other hand, lower values of dispersion will cause severe amplitude fluctuations to develop. As a result, the spectrum becomes chaotic, and no comb forms (Fig. \ref{fig:theory}A right). When the gain is lowered but curvature is left on, the laser instead enters a CW mode, as
the diffusion counteracts the cross-steepening nonlinearity responsible for FM comb formation. Not only does gain curvature limit the bandwidth, but it also limits the dynamic range over which combs can form. The effects of gain curvature can have deleterious effects even if a comb forms. This can be dramatically observed in the case of double-peaked gain media (Fig. \ref{fig:theory}D), which are frequently encountered in THz QCLs \cite{kumarCoherenceResonanttunnelingTransport2009,burghoffUnravelingOriginFrequency2020}. Even just 10\% of gain modulation can lead to combs whose amplitude varies by \textit{two orders of magnitude}. This is also in contrast to pseudorandom states, whose bandwidths were found to scale with $\Omega^{2/3}$ \cite{khurginAnalyticalExpressionWidth2020}.

\begin{figure}
\centering
\includegraphics{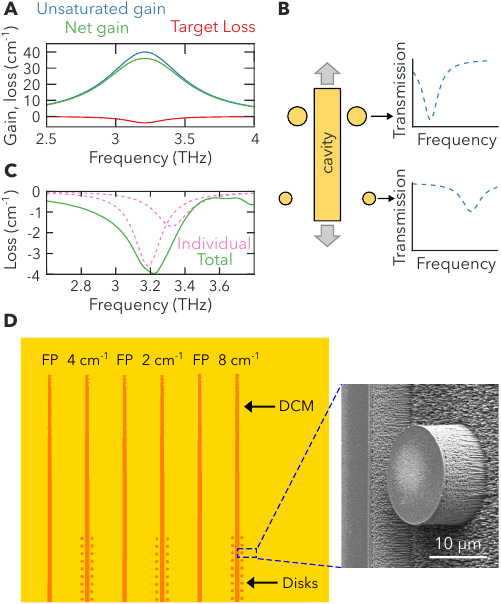}
\caption{\textbf{Loss shaping of THz QCL combs.} \textbf{(A)} Strategy to reduce gain curvature. The goal is to introduce loss centering at the same frequency as that of the unsaturated gain so that the curvature of the net gain is reduced. \textbf{(B)} Unit cell of the design, in which multiple disk pairs are evanescently coupled to the laser. By adjusting the radius and coupling distance of the disks, the losses can be precisely controlled. \textbf{(C)} Individual and total
disk losses for a device with 4 cm$^{-1}$ of peak loss (see ``Methods" for device parameters). Multiple disk pairs can be used per unit cell to achieve broader loss spectra and to finely tune the profile. \textbf{(D)} Structures that were designed to demonstrate the efficacy of the anti-diffusive loss shaping. FP lasers were interspersed with shaping structures designed to add 2, 4, and 8 $\si{\cm^{-1}}$ of peak loss.}
\label{fig:fab}
\end{figure}

It is worth pointing out that for typical lasers, the constant of proportionality relating the comb bandwidth to full-width half maximum (FWHM) gain bandwidth is somwehat small (0.15-0.3). Therefore, while the ultimate limit is indeed proportional, there is significant room for improvement, as much more of the laser's gain bandwidth is above threshold. To demonstrate that bandwidth can be enhanced and that the deleterious effects of gain curvature can be removed, we engineered anti-diffusive loss structures that introduce resonant loss with the same center frequency as that of the gain (Fig. \ref{fig:fab}A). These loss structures are anti-diffusive because  their effect can be viewed as adding negative diffusion to Eq. \ref{nlse}. Since the loss adds a \textit{positive} $\frac{\partial^2g}{\partial\omega^2}$, the net gain becomes flat.

For this work, we demonstrated our concept on THz QCL combs made in the metal-metal waveguide platform. Specifically, we introduce small disks on both sides of a FP  laser cavity, covering almost the entire length of the cavity. Each of the disks is, in fact, the QCL gain medium in a metal-metal waveguide. Due to the proximity of these disks to the cavity, light couples into these disks, which results in cavity loss. The loss has resonance characteristics, which can be controlled by varying the disk parameters. Figure \ref{fig:fab}B shows a schematic view of a unit cell of
our design, in which the whole structure is obtained by simply replicating the unit cell periodically. The structure is symmetric with respect to the plane bisecting the FP cavity, which mitigates scattering of the fundamental
lateral mode of the FP cavity into higher-order asymmetric modes.

Therefore, through an appropriate design, it is possible to reduce the net gain curvature of the overall gain and increase comb bandwidth. In contrast to the heterogeneous gain medium concept, our strategy is less susceptible to material growth uncertainties, as we are able to tailor our cavities electromagnetically and even for a particular wafer. It can also compensate for unwanted curvature in homogeneous wafers (such as double-peaked media).

\section{\label{performance}Results and Discussion}
The active region of the QCL we chose for our work uses a regrown version of the same design used in Ref. \cite{curwenBroadbandContinuousSinglemode2019}. This is a photon-phonon \cite{amantiBoundtocontinuumTerahertzQuantum2009,curwenBroadbandMetasurfaceDesign2020} design that has low threshold and broadband operation, which when processed into tunable VECSELs lased over a range of 880 GHz \cite{curwenBroadbandContinuousSinglemode2019}. However, since this operation did not all occur at the same bias, it should be considered an upper bound for the maximum possible bandwidth of the structure above threshold. Furthermore, our regrown wafer never demonstrated any evidence of gain above threshold beyond the 3--3.7 THz range. To design our shaping structures, we assumed that the unsaturated gain was a Lorentzian with a peak value of 40 $\si{\cm^{-1}}$ and had an FWHM of 0.67 THz \cite{burghoffTerahertzPulseEmitter2011}.

For most of these structures, we seek to add a relatively small amount of engineered loss, ranging from 2 to 8 $\si{\cm^{-1}}$. Though the overall effect on the gain profile is relatively minor (Fig. \ref{fig:fab}A), this can be sufficient to eliminate the curvature around the gain peak. Figure \ref{fig:fab}C shows the individual and total loss obtained by simulating a structure that has two pairs of disks per unit cell (see Fig. S2 for the simulation of other designs). Since the FWHM of the total loss depends on the radii of the disks used, one might include more disk pairs of varying radii to increase the width of the total loss. Moreover, by varying the distance between the disks and the laser cavity (i.e., the coupling distance), losses of different amplitude can be introduced. 
To verify the efficacy of our approach, we designed a variety of loss shapers (single- and two-disk-pair devices with 2, 4, and 8 $\si{\cm^{-1}}$ of peak loss), and these were interspersed with FPs (Fig. \ref{fig:fab}D). All of these structures were designed with double-chirped mirrors (DCMs) to eliminate dispersion  (designed to compensate for dispersion of 0.1 $\si{ps^2/mm}$) \cite{burghoffTerahertzLaserFrequency2014}. The details about the devices fabricated are provided in Supplementary Materials (see ``Device parameters").
The fabrication of these devices was done following a standard metal-metal waveguide process.

\begin{figure}
\centering
\includegraphics{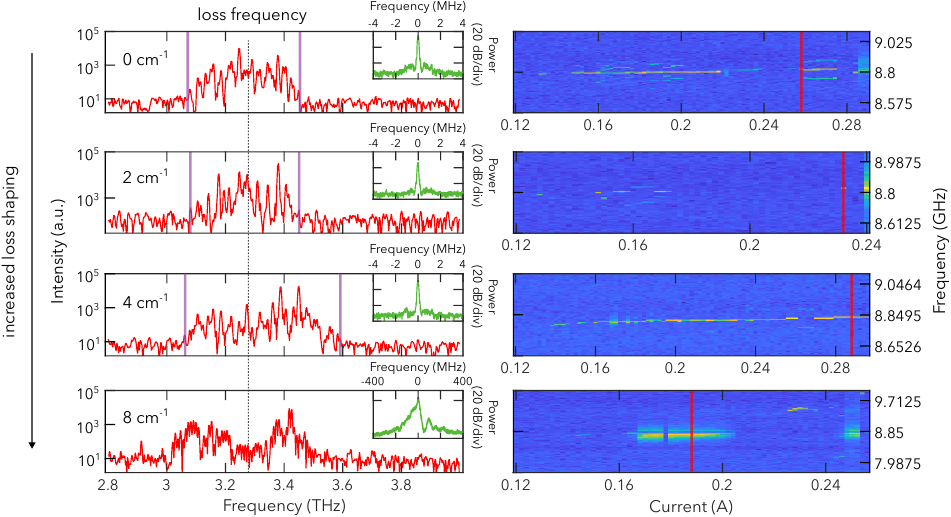}
\caption{\textbf{Spectra and beatnote maps for the loss-shaped devices.} Effect of anti-diffusive loss shaping on THz QCL combs, with spectra shown on the left, corresponding beatnotes in the insets, and beatnote maps on the right. Starting from the top, the structures possess a single disk pair per unit cell and have 0, 2, 4, and 8 $\si{\cm^{-1}}$ of loss shaping. The spectrum and beatnote map improve significantly for the device with 4 $\si{\cm^{-1}}$ of shaping, but the final one adds too much loss and the spectrum bifurcates, and no stable comb regime was found. All the measurements were performed at a temperature of 32 K.}
\label{fig:curvature_series}
\end{figure}

In Fig. \ref{fig:curvature_series}, we show the spectra and beatnote map for an FP (indicated by 0 $\si{\cm^{-1}}$) and three single-disk-pair devices with 2, 4, and 8 $\si{\cm^{-1}}$ of resonant loss, respectively (see Fig. S3 in Supplementary Materials for voltage-light-current density characteristics). The spectra shown here are the broadest spectra measured with narrow beatnotes for the respective devices, and the corresponding current biases are indicated on the map (vertical red line). The spectra were measured with a nitrogen-purged FTIR and pyroelectric detector, so atmospheric absorption lines are apparent and prevent measurement above 3.6 THz, where a deep absorption line is present. The devices tested are 3.9 mm long, 20 $\si{\mu m}$ wide, and contain DCMs with the same parameters. The beatnote map shows that the FP and devices with 2 and 4 $\si{\cm^{-1}}$ losses operate as frequency combs for certain bias ranges. Comparing the spectra for these devices, we find that the comb bandwidth for the 4-$\si{\cm^{-1}}$-loss device is the broadest, about 200 GHz (30\%) broader than that for the device without any disks. In addition, the beatnote map of this device is by far the cleanest, operating as a comb essentially across the full dynamic range of the laser (without any additional tuning). A minor dip in the spectrum appears at the peak-loss frequency ($\sim$3.28 THz), but the overall flatness, bandwidth, and dynamic range make this comb much better than that of the reference FP. The spectrum of the 8-$\si{\cm^{-1}}$ device is too strongly affected by the addition of loss, and as a result, it has a large hole at the peak-loss frequency. Additionally, it no longer possessed any stable comb regimes, only incoherent multimode regimes.

\begin{figure}
\centering
\includegraphics{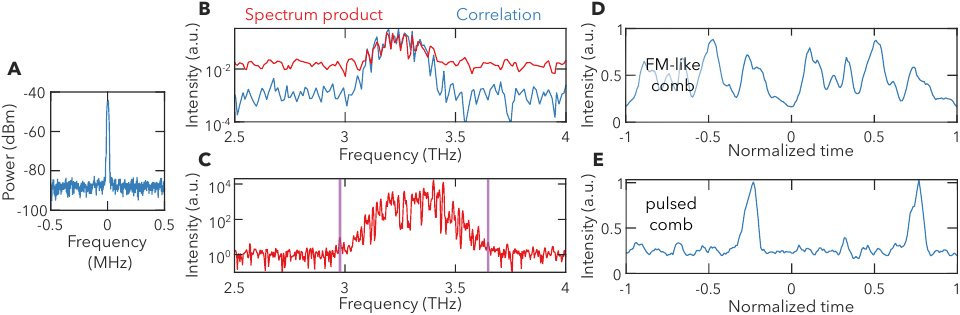}
	\caption{\textbf{Temporal characterization of the 4-cm$^{-1}$-loss device.} \textbf{(A)} Optical beatnote as measured on a room-temperature Schottky mixer in the broadest-bandwidth comb state. \textbf{(B)} SWIFTS coherence measurement using the self-referenced scheme \cite{burghoffEvaluatingCoherenceTimedomain2015} at the broadest-bandwidth bias, demonstrating coherence across the full range over which the spectrum product is above noise. \textbf{(C)}  Spectrum measured using a high-dynamic-range superconducting bolometer, demonstrating 700 GHz of comb bandwidth. \textbf{(D)} SWIFTS intensity for the broadest-bandwidth comb, in which the intensity is only weakly modulated (FM-type). \textbf{(E)} SWIFTS intensity profile for a comb in a pulsed comb regime, with pulses approximately 4 times the height of the FM portion of the wave. All the measurements were performed at 32 K.}
 \label{fig:swifts}
\end{figure}

In order to more fully characterize these devices and to understand their dynamics, coherence and temporal measurements were performed using SWIFTS \cite{burghoffTerahertzLaserFrequency2014,burghoffEvaluatingCoherenceTimedomain2015,hanSensitivitySWIFTSpectroscopy2020}. A room-temperature Schottky mixer (WR0.22FM from Virginia Diodes, with response up to 5 THz) was used to detect the optical beatnote, and high-signal-to-noise ratio (SNR) beatnotes (40 dB) could be achieved (Fig. \ref{fig:swifts}A). Despite the high SNR in the intermediate frequency (IF) chain, the DC-coupled monitor signal is not intended to be low-noise and had limited SNR. Despite this, in the broadest-bandwidth comb regime, the device has a coherence spectrum that matches the normal spectrum product well at the frequency ranges where the spectrum product is above the noise floor (Fig. \ref{fig:swifts}B). To verify the ultimate bandwidth of the comb and compare with the best theoretical performance, the spectrum was measured with a higher-dynamic-range superconducting bolometer (Fig. \ref{fig:swifts}C). The spectrum is flat at the top and has SWIFTS temporal traces consistent with FM modes of operation. Although the blue side of the spectrum is difficult to precisely confirm owing to the strong atmospheric absorption around 3.6 THz, our results establish a comb bandwidth of at least 700 GHz, which is approximately 80\% of the 880 GHz range over which a nominally identical wafer was able to lase over \cite{curwenBroadbandContinuousSinglemode2019}. While this comparison is not ideal---even nominally identical wafers can be different, and a single-mode laser has the benefit of bias tuning---but this illustrates that  the loss shaping can result in combs that are nearly as broad as the bandwidth of the gain spectrum above threshold. Other regrowths achieved even broader lasing \cite{curwenBroadbandMetasurfaceDesign2020}, but this regrowth of the gain medium never showed any evidence of lasing beyond the 3--3.7 THz range, at any bias. 

Interestingly, the 4-cm$^{-1}$ device possesses unexpected pulsed modes of operation as well. While the spectra exhibit primarily FM-like characteristics at most biases (see Fig. \ref{fig:swifts}D and Fig. S4 in Supplementary Materials), clear pulses form instead when the comb is biased very near an unstable regime (see Fig. \ref{fig:swifts}E). 
The pulses are strikingly large---approximately 4 times larger than the FM portion of the wave---and would generally be considered more characteristic of solitons than of extendons. Though QCLs do not form pulses readily, THz QCLs are more amenable to pulse formation on account of their longer gain recovery times \cite{riccardiShortPulseGeneration2023,michelettiTerahertzOpticalSolitons2023,wangUltrafastResponseHarmonic2020,baconGainRecoveryTime2016,barbieriCoherentSamplingActive2011}. Although the limited SNR  makes precise evaluation of the origin challenging, it likely stems from the boundary pulse occasionally observed in FM combs
\cite{singletonEvidenceLinearChirp2018,sterczewskiFrequencymodulatedDiodeLaser2020}. This boundary pulse appears in simulations as well but is usually not so pronounced (e.g., see the power in Fig. \ref{fig:theory}E). 
While a full analytical theory that exactly captures the effect of the boundary pulse has not yet been developed, it is known from numerics that multimode behavior arises at the boundary and that the boundary pulse is largest at the edge of stability (which is the case here).

Going forward, our results imply new degrees of freedom in the engineering of active cavity combs, especially QCLs combs. Up until now, the development of novel comb states has primarily focused on the dispersion degree of freedom, and to the extent that gain has been engineered, it has been through the creation of broadband heterogeneous designs. While heterogeneous designs will likely be critical for achieving octave-spanning combs, achieving sufficiently flat gain spectra through MBE growth alone would be a monumental task. The very act of combining multiple stacks naturally leads to gain variations, and even minor (few $\si{\cm^{-1}}$) variations can lead to enormous changes in performance. However, loss shaping allows for the possibility of trimming these variations, not just using a fixed fabricated design but even dynamically. Adding bias to these structures would allow the resonant loss to become resonant gain, giving even finer control over the comb spectrum. Furthermore, our strategy is compatible with every type of cavity, including enhanced FPs \cite{senica_frequency-modulated_2023} and rings \cite{mengDissipativeKerrSolitons2022,kazakov_active_2024,opacak_nozakibekki_2024,khan_frequency_2023}.

\section{\label{conclusion}Conclusion}
In conclusion, we have both theoretically and experimentally demonstrated the pivotal role gain curvature plays in shaping FM combs. Using a mean-field theory, we showed that the maximum FM comb bandwidth in general varies linearly with effective gain bandwidth. Furthermore, we showed how it can be modified.
By introducing the concept of anti-diffusive loss shaping---employing resonant structures closely coupled to a laser cavity---and applying this approach to THz QCLs, we eliminated gain curvature and realized comb bandwidths near the intrinsic limit of the medium (80\% of the maximum lasing bandwidth of a single-mode laser). We verified the coherence and the FM nature of the combs using  SWIFTS. 
This research unveils the enormous potential harbored by the simultaneous engineering of both dispersion and diffusion, and the introduction of loss shaping in active cavities paves the way for more robust, versatile, and advanced chip-scale frequency comb systems.

\section*{Methods}
\subsection*{\label{fab}Device fabrication and design}
First, Ta/Au (10 nm/250 nm) was deposited on both the MBE sample ($\sim$645 $\mu$m thick) and a heavily doped n-type
GaAs sample (receptor). The two samples were then bonded via thermo-compression at \SI{300}{\celsius} \cite{williams4THzQuantumCascade2003}. Next, the MBE sample was mechanically lapped and wet-etched, exposing the top layer of the active region. Then, the device patterns were transferred onto the sample by electron beam lithography, and Ti/Au/Ni (10 nm/160 nm/230 nm) was deposited to define the top metal layer. Using this layer as a mask, dry etching (BCl$_{3}$/Cl$_{2}$/Ar) was then performed to define the devices. Finally, the sample was thinned down to $\sim$200 $\mu$m by mechanical lapping of the backside, and Ti/Au (20 nm/120 nm) was deposited on it to facilitate copper-mounting of the devices.

Multiple varieties of loss shapers were designed and tested. The separation between the disks was chosen in a way so that the interference between the disk pairs is minimized. For the multi-disk shaper illustrated in Fig. \ref{fig:fab}C, we chose the width of the underlying FP cavity to be 20 $\mu$m and the radii of the disks to be 8.29 and 8.67 $\mu$m. The disks were placed 19.5 $\mu$m apart from each other and 7.15 and 6.3 $\mu$m away from the cavity. 

\subsection*{\label{swifts}SWIFT spectroscopy}
A self-referenced scheme was used to perform SWIFTS. THz emission from the QCL was collimated through an \textit{f}/3 off-axis parabolic mirror and passed through a custom-built FTIR. Roof mirrors were used to build the FTIR to minimize reflection back to the device. A Virginia Diodes Schottky mixer (WR0.22FM) with a frequency response of up to 5 THz was used to detect the FTIR output. The mixer monitor port was connected to a lock-in amplifier for normal interferogram. The signal from the IF port was amplified using two low-noise amplifiers (LNAs) and passed through the LO port of a 9 GHz I/Q demodulator, DC2645A. The RF signal from the QCL was extracted using a bias tee, amplified using two LNAs, and fed into the RF port of the demodulator. The I and Q outputs from the demodulator were collected using two lock-in amplifiers. All the lock-ins were set to have a similar time constant and phase. The beatnote from the QCL was observed using a spectrum analyzer connected to a nearby unbiased QCL.

\bibliography{burghoff_library2}
\bibliographystyle{ScienceAdvances}

\clearpage

\end{document}